%% file: main.tex
\documentclass[10pt, conference]{IEEEtran}

\usepackage[utf8]{inputenc}      

\usepackage{microtype}
\usepackage{amsmath}
\usepackage{hyperref}
\usepackage{graphicx}
\usepackage{adjustbox}
\usepackage{booktabs}
\usepackage{multirow}
\usepackage{multicol}

\begin{document}

\title{Estimating \& Mitigating the Impact of Acoustic Environments on Machine-to-Machine Signalling}

       
\author{\IEEEauthorblockN{Amogh~Matt}
\IEEEauthorblockA{\textit{Machine Listening Lab, Centre for Digital Music} \\
\textit{Queen Mary University of London}\\
London, England \\
amogh.matt@gmail.com}
\and
\IEEEauthorblockN{Dan~Stowell}
\IEEEauthorblockA{\textit{Machine Listening Lab, Centre for Digital Music} \\
\textit{Queen Mary University of London}\\
London, England \\
dan.stowell@qmul.ac.uk}
}






\maketitle

\input{abstract.tex}

%
\IEEEpeerreviewmaketitle

\input{introduction.tex}
\section{Machine-to-Machine Signals}
For the purpose of this research we focus on an M2M audio codec developed by Chirp, a London based Data-over-Sound communication company. Chirp (not to be confused with the conventional signal processing term: chirp (sweep) tone) enables devices to share data among them via sound. The transfer of data is done via a dynamically generated monophonic/polyphonic audio signal referred to as a packet. Frequency-shift-keying (FSK), a popular DSP modulation technique, is used to encode the data. The encoded data can then be broadcast by a speaker and received by a microphone. The decoding of the packet is done via demodulation. A Chirp signal (comprised of packets) is designed to be robust over distances of several metres, in noisy, everyday environments.
\begin{figure}[ht]
  \centering
  \includegraphics[page=1,width=0.99\linewidth]{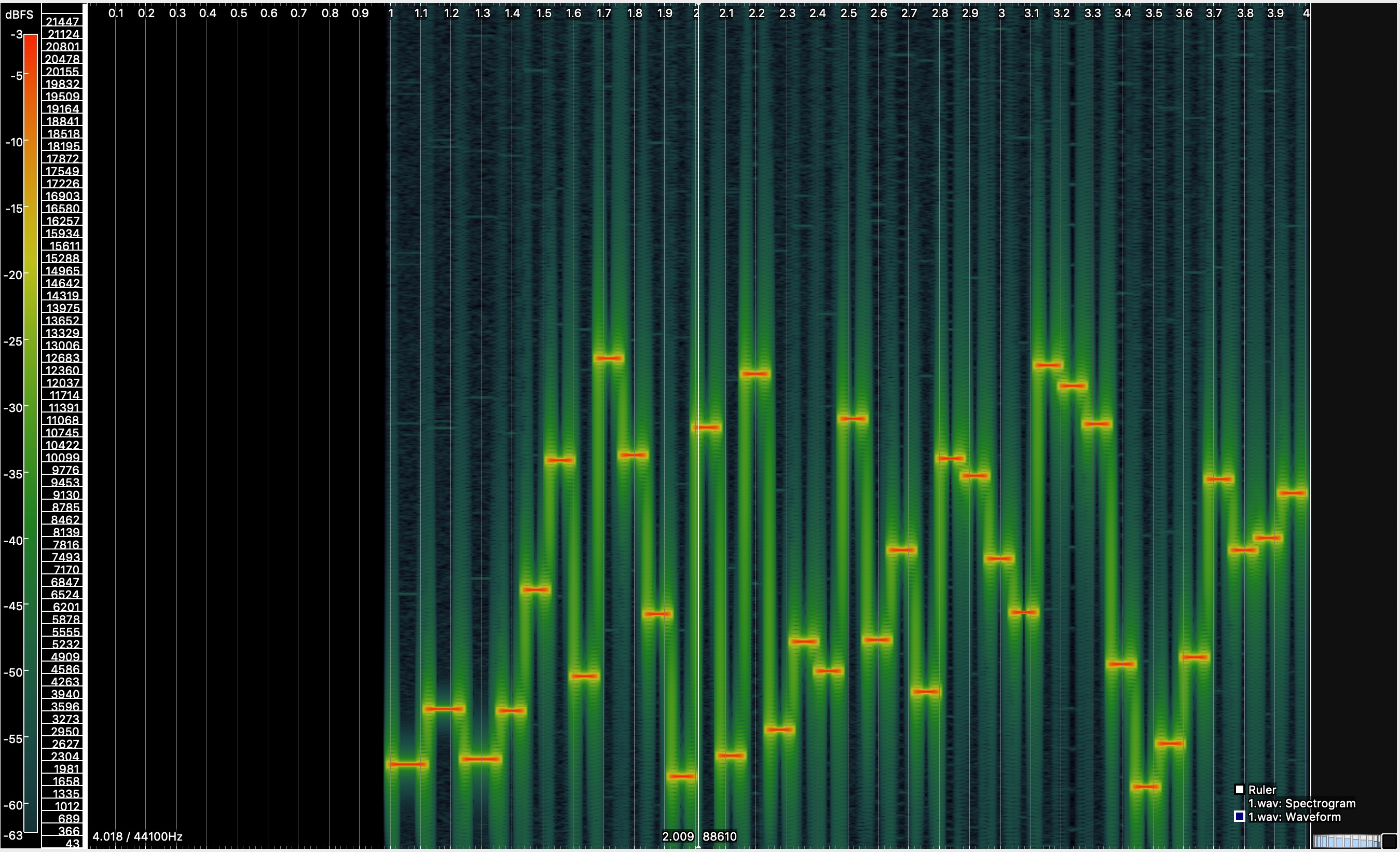}
  \caption{Spectrogram of an anechoic Chirp signal.}
  \label{fig:rawchirp}
\end{figure}
A single packet comprises of preamble tones, a payload length tone, the payload tones and error correction tones. Figure 1 shows the spectrogram of a Chirp packet with labels indicating these components. The sequence of frequencies corresponding to each category is dynamically determined by the data being transmitted. The preamble tones are fixed for all the packets utilising a particular Chirp protocol. The audible protocol generates a Chirp signal within the 1.7kHz to 10.5kHz band and the inaudible protocol utilises the ultrasonic frequency band (18-20 kHz). Each symbol byte of data is then converted into a monophonic tone with a mapped frequency using FSK. Chirp packets employ Reed-Solomon (RS) error correction algorithm. The RS encoder takes in the payload as input and adds extra redundant bits. The decoder processes the body (payload + RS codes) and attempts to correct the errors and erasures and recover the original data.

There are many degradations along the signal path that impair the ability of the decoder to decipher the transmission. These include background noise, distortions, reverberation etc. The effect of reverberation on a Chirp signal is to cause it to sound distant and spectrally modified. This reduces the intelligibility of a Chirp signal  \cite{Naylor:2010:SD:1892051}. Figure 2 is a spectrogram illustrating the change in the Chirp signal when convolved with the impulse response of a room impulse response (RIR) having $RT_{60}$ of 1.2 seconds. 
Looking at the spectrogram, we notice the presence of reverberation tails for each note. It is evident that the energy decay of preceding tones interferes with the subsequent one, leading to a strong risk of mis-detections. 
Hence there is a need for a dereverberation algorithm that functions across many different environments. 
\begin{figure}[t]
  \centering
  \includegraphics[page=1,width=0.99\linewidth]{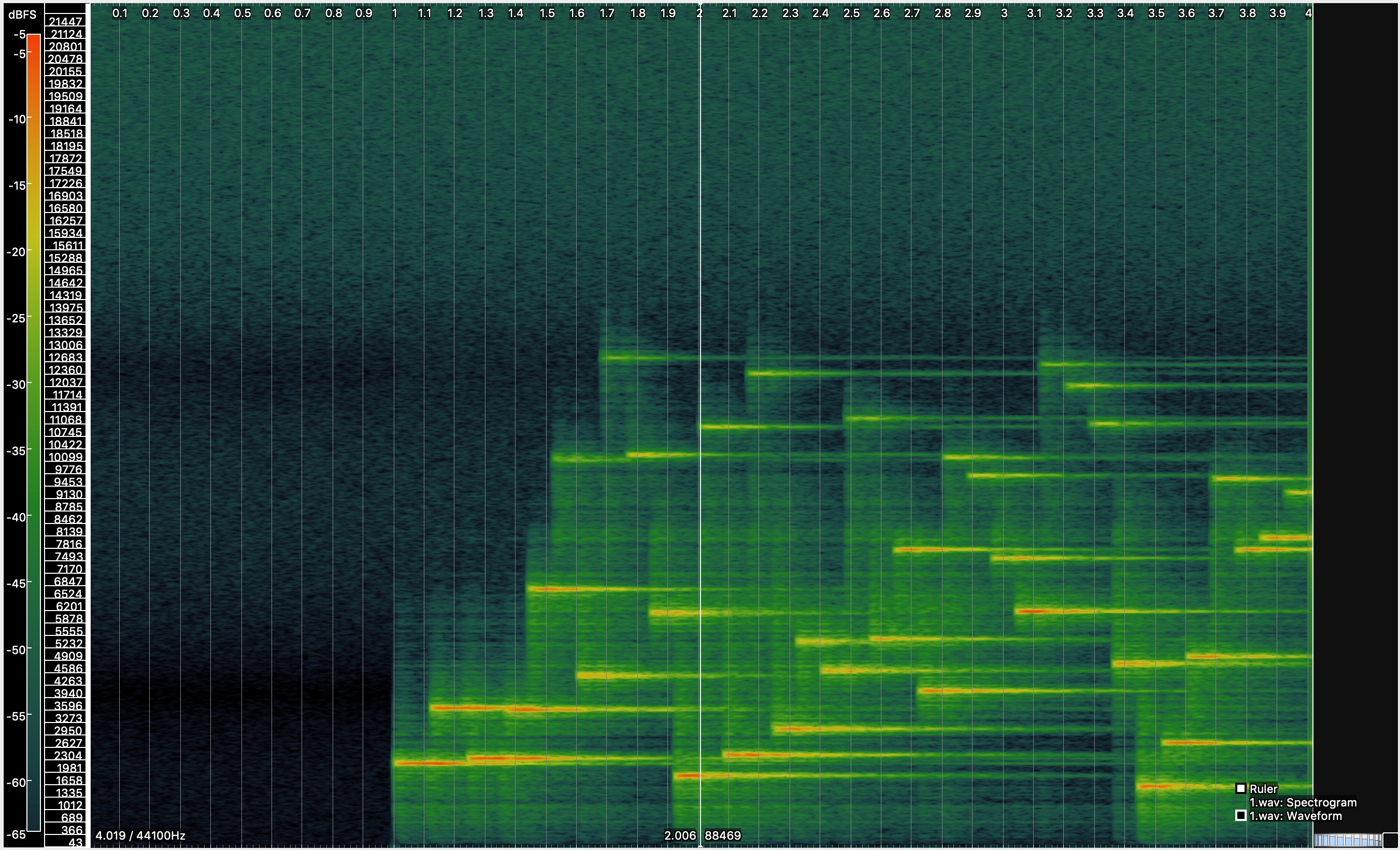}
  \caption{Spectrogram of a reverberant Chirp signal.}
  \label{fig:reverberantchirp}
\end{figure}
\section{Proposed dereverberation algorithm}
The target applications for the proposed algorithm are often single-channel audio, which means existing popular multichannel dereverberation techniques cannot be used. Also, as we do not have any prior knowledge of the RIR we cannot perform reverberation cancellation. Hence, we focus on the dereverberation techniques that suppress the late reflections.
A popular single channel reverberation suppression algorithm is spectral subtraction  \cite{habets}.
Figure 3 shows an overview of the spectral subtraction algorithm. The main goal of this dereverberation method is to find an appropriate gain function. The gain function is tuned to minimise the error between the source signal and the reverberant signal. This is done by estimating the power spectral densities (PSD) of the reverberant signal.

\begin{figure}[ht]
  \centering
  \includegraphics[page=1,width=0.99\linewidth,clip,trim=0mm 75mm 0mm 60mm]{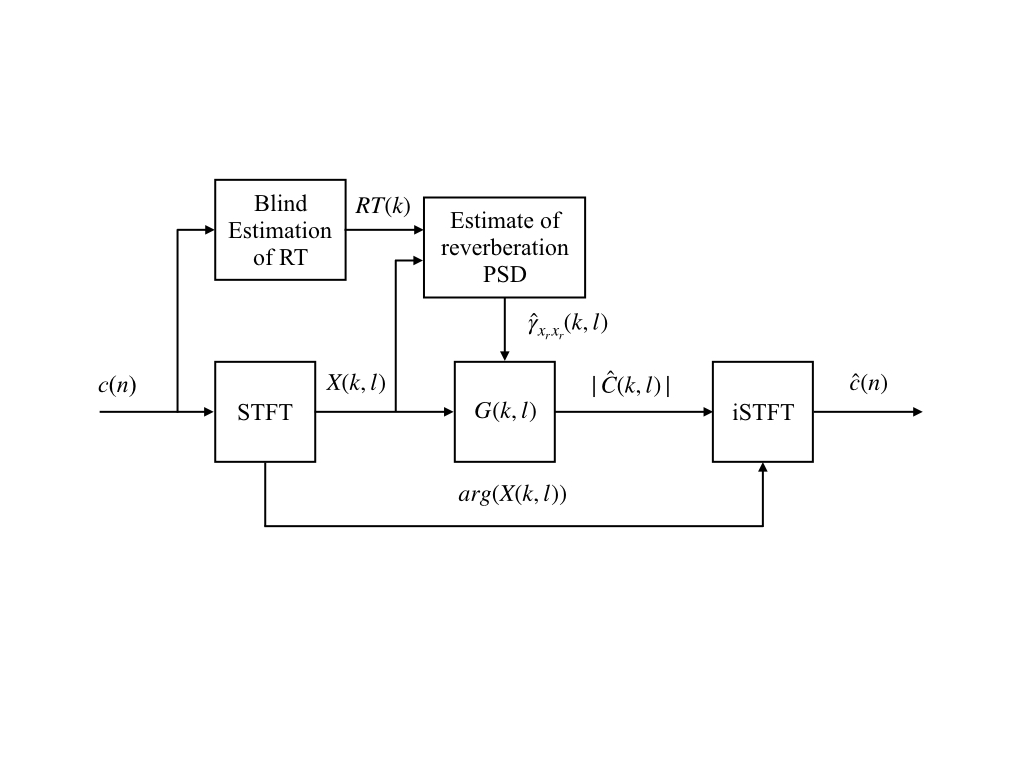}
  \caption{Schematic overview of the algorithm.}
  \label{fig:algorithm}
\end{figure}

In general, there are 3 main stages of the algorithm. The first estimates the RT of the reverberated Chirp signal, the second stage computes the reverberant PSD. This is used to obtain the real valued gain function G(k,l). In the last stage, the gain is multiplied with the reverberant input to obtain the dereverberated output short-time spectrum C(k,l). An inverse short-time Fourier transformation synthesises  the dereverberated signal. 

\subsection{Joint estimation of EDC and RT60}
Polack models the RIR statistically as an exponentially decaying white Gaussian noise which is based on the reverberation time $RT_{60}$  \cite{polack1988transmission}. The energy decay of the RIR is given by (1). 

\begin{equation}
EDC(t) = \int_{\tau=t}^\infty h^2(\tau)d\tau					
\end{equation}

Since the parameters of this model are frequency independent, we can extend this model of (1) into STFT discrete time and assume that the decaying model is valid in each frequency bin. For frequency bin k this results in 

\begin{equation}
EDC_k(t) = \int_{\tau=t}^Th_k^2(\tau)d\tau					
\end{equation}

Where $h_k^2(t$) is the energy envelope of frequency bin $k$. Figure 4 shows the log-energy envelope of a subband of a reverberant Chirp signal obtained through STFT. It is observed that after the initial burst of energy (corresponding to the Chirp tone) the energy decays in correlation with its RIR. 

A peak picking algorithm is used to obtain the time of the energy peak of the subband. The peak corresponds to the attack portion of the Chirp tone envelope. The start of the EDC is offset from the peak by a constant value as an estimated time of the end of the Chirp tone. Most subbands decay to the noise floor and limit the dynamic range to less than 60 dB. To account for this, the decay rate is estimated by a linear least-squares regression of the measured decay curve from a level 5 dB below the initial level to 35 dB. The linear least-square regression is performed over the cumulative sum of the normalised energy envelope values corresponding to -5 dB and -35 dB. Figure 4 also shows the fit of the regression over the log-energy envelope of the subband. The value of $RT_{60,k}$ is obtained by the x-intercept corresponding to 60 dB. 

\begin{figure}[ht]
  \centering
  \includegraphics[page=1,width=0.99\linewidth,clip,trim=0mm 0mm 0mm 8mm]{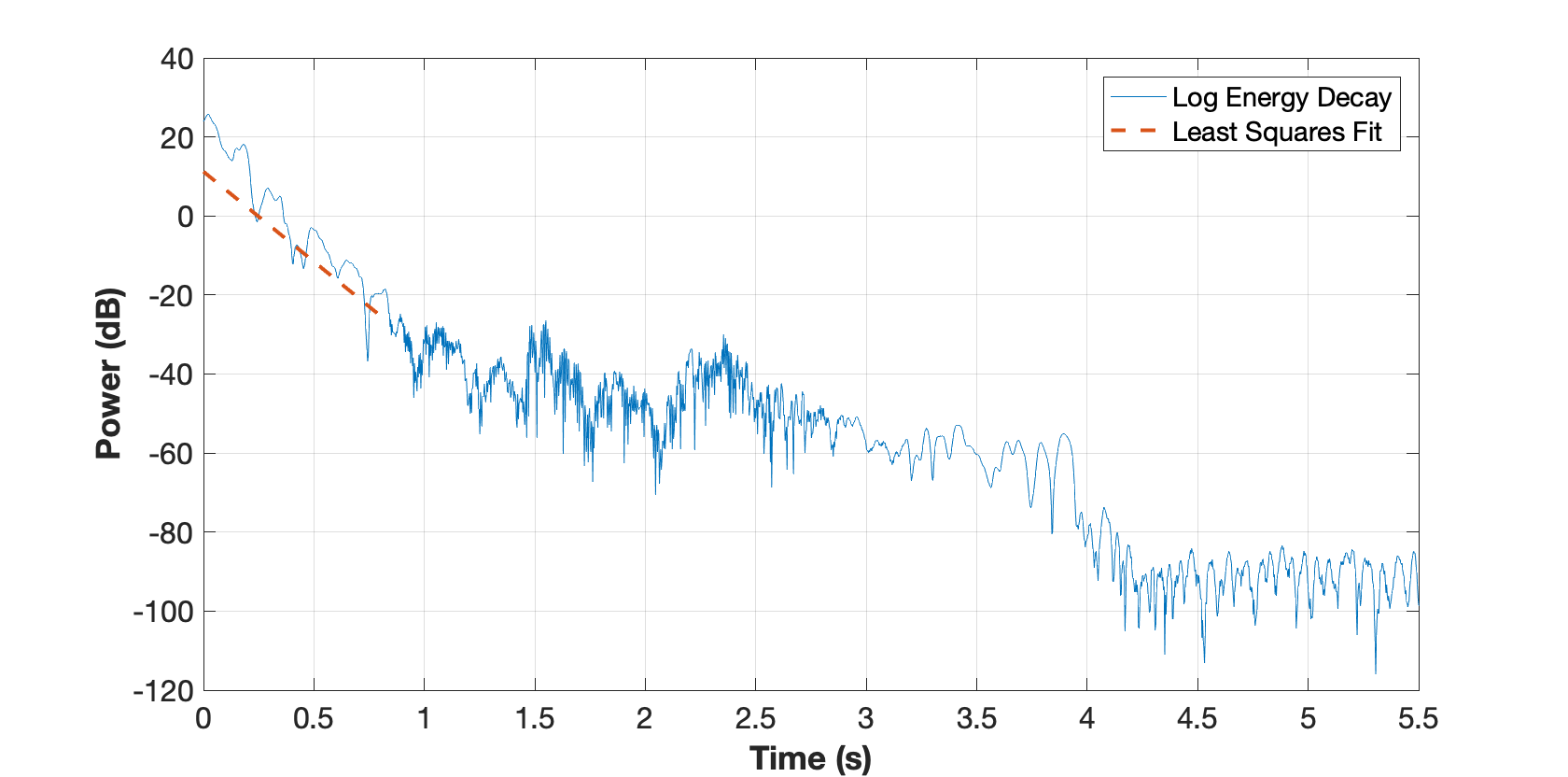}
  \caption{Least-squares regression fit over the log-energy decay.}
  \label{fig:lsfit}
\end{figure}

Subbands are thresholded before processing to ignore bands which do not contain any meaningful information. The overall $RT_{60}$ is calculated by averaging the nonzero $RT_{60,k}$ estimates

\begin{equation}
RT_{60}= \sum_{k=1}^K \frac{RT_{60,k}}{K} \forall RT_{60,k} \ne 0
\end{equation}

\subsection{Estimating the Reverberation PSD}

A component view of the 'Estimate of Reverberation PSD block' (from Figure 3) is shown in Figure 5. The reverberant portion of the source signal is estimated from the previous frames using the short-term power spectrum of reverberated source. Due to the convolution of the RIR with the Chirp signal, the reverberant Chirp signal exhibit the exponential decaying envelope found in the RIR. The Chirp signals are stationary over periods of time that are short compared to the $RT_{60}$. If $T_c$ be the period where the Chirp signal is stationary. The short-term power spectral densities are given as the auto-correlation at time l by 

\begin{equation}
\gamma_{xx}(k,l) = \gamma_{x_r x_r}(k,l) + \gamma_{x_d x_d}(k,l) 
\end{equation}

\begin{figure}[th]
  \centering
  \includegraphics[page=1,width=0.99\linewidth,clip,trim=0mm 85mm 0mm 75mm]{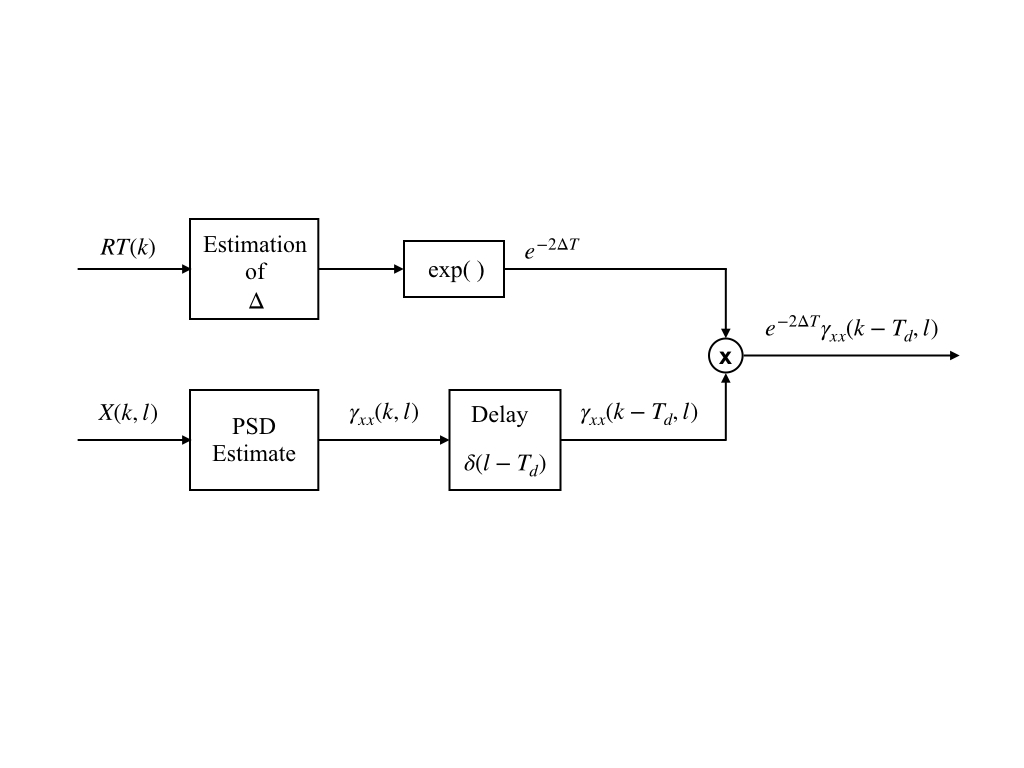}
  \caption{Schematic overview of the Reverberation PSD Estimation.}
  \label{fig:reverberation_psd}
\end{figure}
where $x_d(t)$ and $x_r(t)$ are the result from the convolution of the Chirp signal $c(t)$ with the early (direct) part $h_d(t)$ and late reflection part $h_r(t)$ of the RIR. The short-term PSD of the late reverberant signal from (4) is estimated using a delayed and attenuated version of the short-term PSD.

\begin{equation}
\gamma_{x_r x_r}(k,l) = e^{-2\Delta T}\gamma_{xx}(k,l-T)
\end{equation}
where, $Tc \leq T  \ll RT_{60}$

For our scenario, 80ms was found to be the best resulting value of T and the reverberation estimate starts at frame $l - T$. This estimated reverberation PSD is then subtracted from the reverberated Chirp signal to suppress the reverb tail.

\begin{equation}
|C(k,l)| = X(k,l) - \gamma{x_rx_r}^\frac{1}{2}(k,l) 
\end{equation}

\subsection{Spectral Gain Function}
A common feature of the spectral subtraction technique is that the reverberation suppression process given in (6) can be estimated to be a short-time spectral attenuation factor. In this last block, a real valued spectral attenuation (gain) factor $G(k,l)$ is calculated to suppress the reverberant portions of the Chirp signal $X(k,l)$. 

\begin{equation}
 |C(k,l)| = |X(k,l)| G(k,l)	
\end{equation}

The short-term spectral attenuation factor according to  \cite{habets} is given by
\begin{equation}
G(k,l) = 1 - \frac{1}{\sqrt{SNR_{post}(k, l)}}	
\end{equation}
Where, $SNR_{post}(k,l)$ is the A-posteriori Signal to Noise Ratio which is calculated by 

\begin{equation}
SNR_{post}(k,l) = \frac{|X(k, l)|^2}{\gamma_{x_r x_r}(k, l)}	
\end{equation}

This fails for scenarios where the estimated reverberant noise amplitude is greater than the instantaneous amplitude of the reverberant Chirp spectrum $|X(k,l)|$. Such scenarios lead to a negative estimate for the amplitude of the clean Chirp spectrum $|C(k,l)|$. For these bins, a half-wave rectification is performed and the gain function $G(k,l)$ is set to 0.		
This rectification however leads to a residual noise problem, which is alleviated by performing the following modifications. 
\subsubsection{Averaging the instantaneous SNR during the calculation of the gain.}
The instantaneous SNR is defined by

\begin{equation}
SNR_{inst}(k,l) = SNR_{post}(k,l) - 1
\end{equation}

The averaged SNR is an estimate of the a priori SNR which is calculated using a moving average of $SNR_{inst}$. This reduces the random variations introduced by late reverberation.

\begin{equation}
\begin{aligned}
SNR_{prio}(k,l)= \beta SNR_{prio}(k,l-1) \\ + (1-\beta) H(SNR_{inst}(k,l)) 
\end{aligned}
\end{equation}

Where $H( )$ denotes the half-wave rectification and $\beta$ is set to 0.9.

\subsubsection{Modifying the half-wave rectification to lead to a non-zero threshold rather than nullifying it.}

\begin{equation}
  G(k,l)=\begin{cases}
    1 - \frac{1}{\sqrt{SNR_{post}(k, l)}}, & \text{if $|C(k,l)| \geq \lambda|X(k,l)|$}.\\
    \lambda, & \text{otherwise}.
  \end{cases}
\end{equation}
Where $\lambda$ is chosen as 0.1.

\section{EVALUATION}

\subsection{Evaluation of RT Estimation}
The $RT_{60}$ estimation algorithm was tested on audible Chirp signals convolved with the 590 RIRs from AcouSP Database\cite{acousp}. These consisted of RIRs with $RT_{60}$ ranging from 0.4s to 16s. Figure 6 shows a graph of the known $RT_{60}$ versus the estimated $RT_{60}$ values obtained using this approach. We find overall that the estimates have a mean error of 0.34s and a mean absolute error of 0.38s. 

For real-world scenarios (RT $<$ 2s) the mean error is 0.11s and mean absolute error is 0.19s. 
We use this estimate of RT in the first block of the spectral subtraction based dereverberation method. 

\begin{figure}[ht]
  \centering
  \includegraphics[page=1,width=0.99\linewidth,clip,trim=0mm 0mm 0mm 10mm]{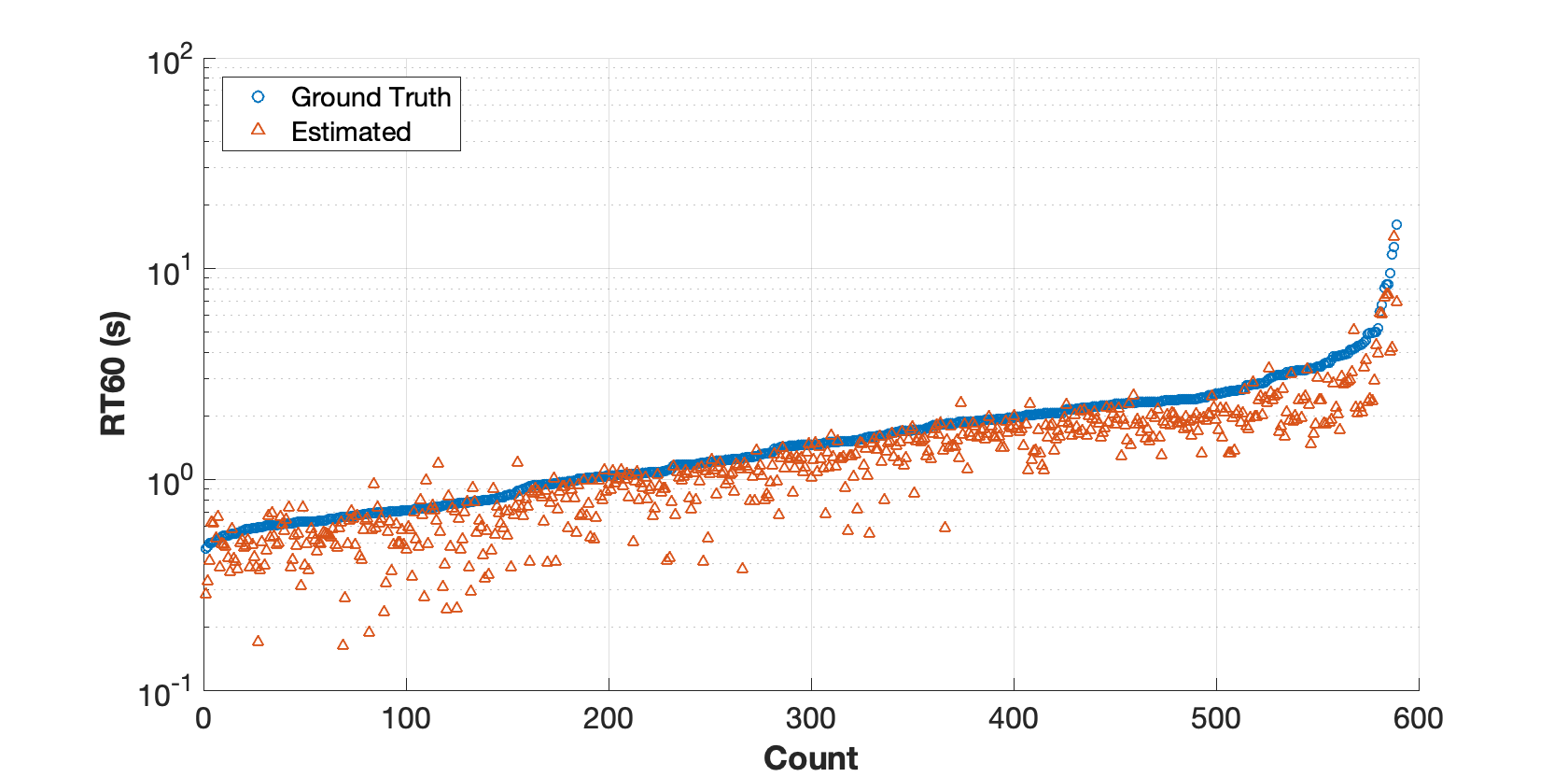}
  \caption{Actual vs estimated reverberation time.}
  \label{fig:rtest}
\end{figure}

\subsection{Evaluation of Dereverberation Algorithm}
\subsubsection{Log Spectral Distortion}
To evaluate the efficiency of the algorithm, the log spectral distortion (LSD) also known as log spectral distance, has been one of the most straightforward and longstanding speech distortion measures \cite{Naylor:2010:SD:1892051}. LSD is the root mean square (RMS) of the difference of the log spectra of the original clean Chirp signal $c(n)$ and the signal $x(n)$ which is the output of the dereveberation model. The LSD is calculated for each time frame on the STFT signals $C(k,l)$ and $X(k,l)$ with $l$ denoting the time frame and $k$ denoting the frequency. The LSD is given by 

\begin{equation}
LSD(l) = (\frac{2}{K} \sum_{k=0}^{\frac{K}{2}-1}|\Gamma\{X(k,l)\}-\Gamma\{S(k,l)\}|^2)^{\frac{1}{2}} \ 
\mathrm{dB}
\end{equation}

where $\Gamma\{X(k,l)\} = max\{20log_{10}(|X(k,l|),\delta\}$ is the log spectrum confined to 50 dB dynamic range, and $\delta = max_{l,k} {20 log_{10}(|X(k,l|)} - 50$ and likewise for $\Gamma\{S(k,l)\}$  \cite{habets}. The mean LSD is obtained by averaging (13) over all the frames containing Chirps. 

\subsubsection{Reverberation Reduction}
Reverberation reduction (RR) is another metric that is used to evaluate the quality of the reverberation model \cite{habets}. RR is the ratio of the sum of the log power on subbands without Chirp tones between the reverberant Chirp signal $c_r(n)$ and the resulting Chirp signal output from the dereverberation model x(n). The RR is calculated for each subband on the STFT signals $C_r(k,l)$ and $X(k,l)$ with $l$ denoting the time frame and $k$ denoting the frequency band. 
The RR is given by

\begin{equation}
RR(k) = 10log_{10}(\frac{\sum_{t=silence}|Cr(k,l)|^2}{\sum_{t=silence}|X(k,l)|^2}) \ \mathrm{dB}
\end{equation}

The mean RR is obtained by averaging (14) over all the subbands. The subbands with silence were determined by thresholding. 

\subsubsection{Evaluation based on the Chirp Decoder}
One hundred audible and another 100 inaudible (ultrasonic) Chirp signals with random packets are generated using the Chirp python SDK. Each of these Chirp signals are convolved with the 590 RIRs from AcouSP database \cite{acousp}, resulting in 59000 audible and 59000 inaudible (ultrasonic) Chirp signals. The 118000 reverberant Chirp signals are decoded to obtain the base decoder performance. Next the spectral subtraction algorithm with the parameter estimation implemented in Matlab is applied on the 118000 Chirp signals. The 118000 dereverbed Chirp signals are then passed to the Chirp python SDK to decode.

\subsubsection{Comparison with popular Single Channel Reverberation Suppression Algorithms}
Two Chirp signals (1 audible and 1 inaudible) with a single packet were each convolved with the 590 RIRs, yielding 1180 (590 audible and 590 inaudible) reverberant Chirp signals. 
Due to constraints in computational complexity Spectral subtraction, LP residual cepstrum and the Source Enhancement algorithms (available as spendred.m in the Matlab Voicebox Toolbox) were chosen for comparison.
Each of the dereverberation methods were programmed on Matlab and applied on these reverberant Chirp signals, the resulting waveforms were then decoded using the Chirp decoder.

\section{RESULTS}
\subsubsection{Technical Quality Metrics}
 An STFT with a 46ms Hann window with 93.75\% overlap is used for calculating LSD and RR. 59000 audible Chirp signals (100 for each IR) are used for the evaluation tests. The mean log spectral distances of the reverberant Chirp signals and the dereverbed Chirp signals from the anechoic Chirp signals averaged over each RIR are shown in Figure 7. 
\begin{figure}[ht]
  \centering
  \includegraphics[page=1,width=0.99\linewidth,clip,trim=0mm 0mm 0mm 5mm]{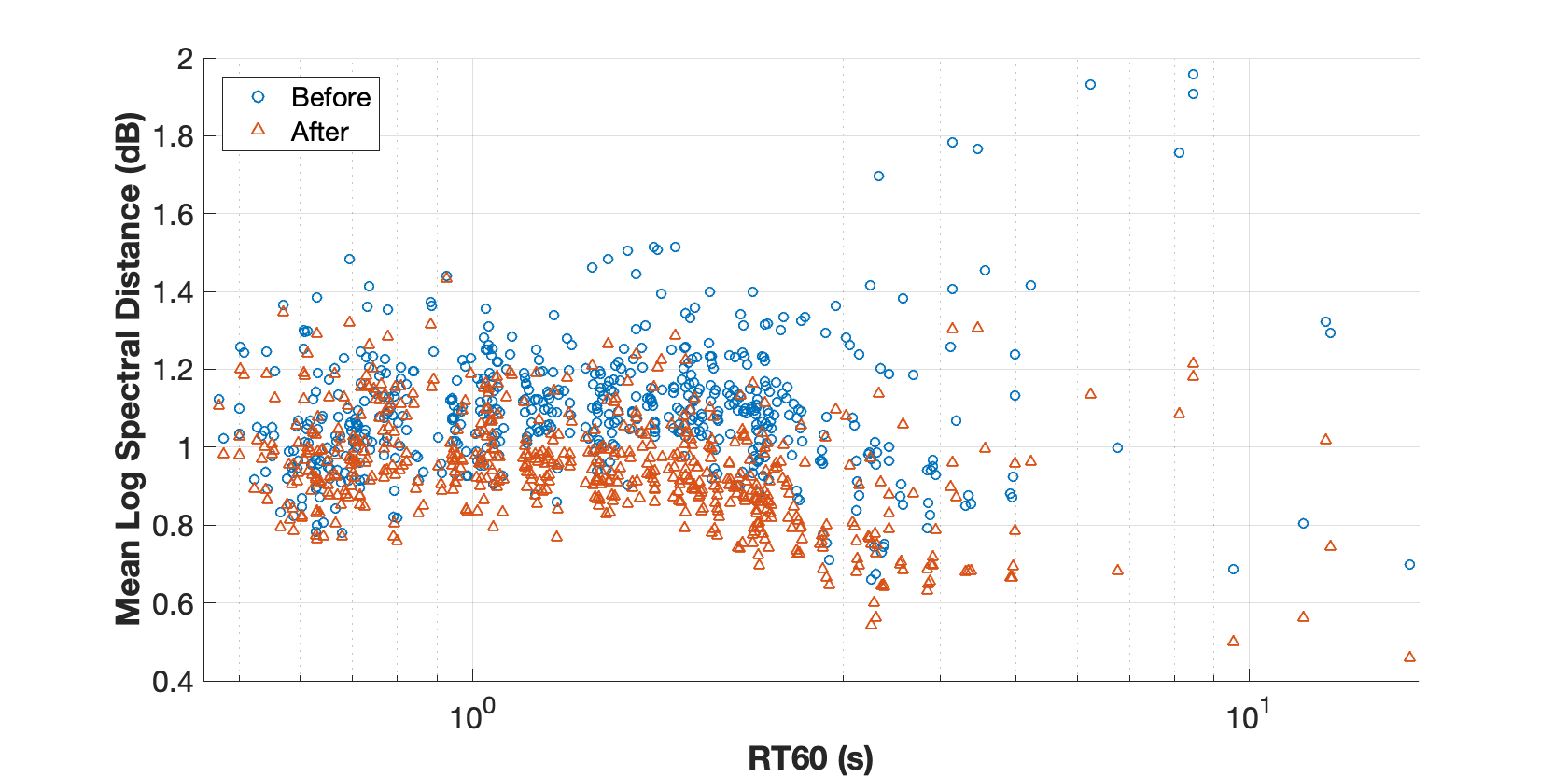}
  \caption{Mean log spectral distance.}
  \label{fig:lsd}
\end{figure}
 It is found that for each RIR, the dereverbed Chirp (shown in red) has a lower LSD than its reverberant counterpart (shown in blue). This signifies that the Chirp signal after dereverberation is closer to the raw Chirp signal than its reverberant counterpart. Figure 8 shows the reverberation reduction in dB for the 590 RIRs between the reverberant Chirp signals and the dereverbed output. A positive correlation is noticed between the RR and the $RT_{60}$. As the RIRs with increasing $RT_{60}$ values have more reverberant parts, an increase in the reverberation portion is noticed. 

\begin{figure}[ht]
  \centering
  \includegraphics[page=1,width=0.99\linewidth,clip,trim=0mm 0mm 0mm 5mm]{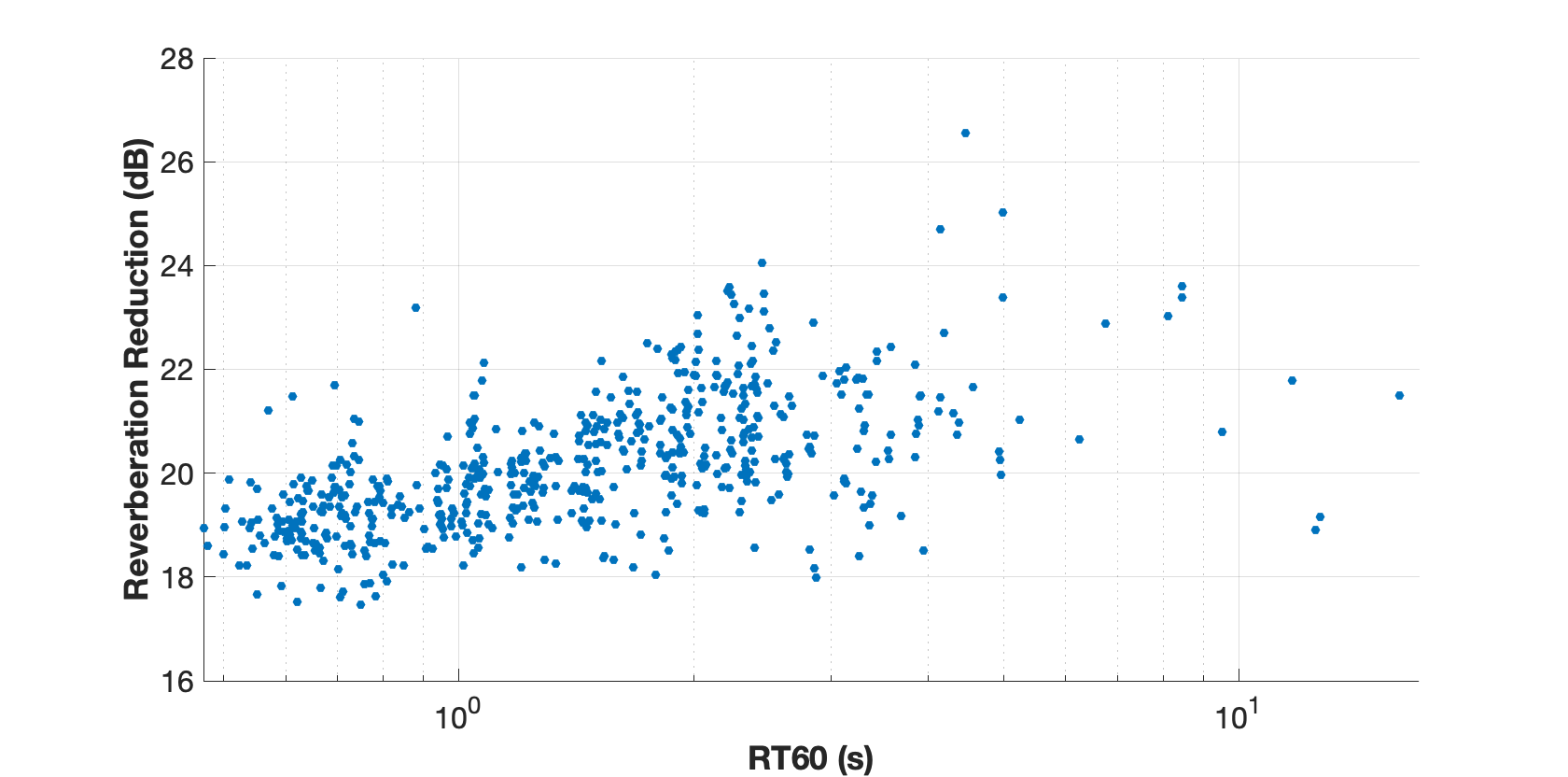}
  \caption{Reverberation Reduction for RIRs with increasing RT.}
  \label{fig:rr}
\end{figure}

\subsubsection{Evaluation based on the Chirp Decoder}
Table I shows the overall results of the dereverberation algorithm. The dereverberation method results in a ~25 percentage points increase in the decode rate of audible Chirp signals and ~6 percentage points increase in inaudible Chirp signals.
In the audible results, it is found that for low $RT_{60}$($<$0.6s) there is a short decrease in the decode rate and an increase in the decode rate for other $RT_{60}$values. For inaudible results, the percentage point increase in result is quite small as the decode rates are significantly high over all. This is expected as the effect of reverberation in the ultrasonic frequency range is comparably less as to that of the audible spectrum \cite{secondscreen}. 

\subsubsection{Comparison with popular Single Channel Reverberation Suppression Algorithms}
Table II shows the performance of the dereverberation algorithms for the audible and inaudible Chirp signals. It was found that our proposed spectral subtraction gave by far the best result, and also the fastest processing time.

\begin{table}[ht]
\label{result_table_1}
\begin{center}
\begin{tabular}{|c|c|c|}
\hline
\textbf{Chirp Signal} & \multicolumn{2}{|c|}{\textbf{Decode Rate (\%)}} \\
\cline{2-3}
 & \textbf{Before} & \textbf{After} \\
\hline
Audible & 54.53 & \textbf{79.78} \\
\hline
Inaudible (Ultrasonic) & 86.15 & \textbf{92.11} \\
\hline
\end{tabular}
\end{center}
\caption{Decode rate of Chirp signals}
\end{table}

\begin{table}[ht]
\label{result_table_2}
\begin{center}
\begin{tabular}{|c|c|c|c|}
\hline
\multirow{2}{*}{\textbf{Algorithm}} & \multicolumn{2}{c|}{\textbf{Decode Rate (\%)}} & \multirow{2}{*}{\textbf{\begin{tabular}[c]{@{}c@{}}Average \\ Processing \\ Time (s) \\ \end{tabular}}} \\ \cline{2-3}
 & \textbf{Audible} & \textbf{\begin{tabular}[c]{@{}c@{}}Inaudible \\ (Ultrasonic)\end{tabular}} &  \\ \hline
\begin{tabular}[c]{@{}c@{}}LP Residual \\ Cepstrum\end{tabular} & 42.54 & 82.03 & 3.20 \\ \hline
\begin{tabular}[c]{@{}c@{}}Source \\ Enhancement\end{tabular} & 44.24 & 81.86 & 17.81 \\ \hline
\begin{tabular}[c]{@{}c@{}}Spectral \\ Subtraction\end{tabular} & \textbf{77.79} & \textbf{87.79} & \textbf{1.76} \\ \hline
\end{tabular}
\end{center}
\caption{Comparison of dereverberation algorithms}
\end{table}

\input{conclusion.tex}
\section{Acknowledgement}
The work of AM was carried out during a placement with \url{chirp.io}.
AM thanks Adib Mehrabi for discussions which informed the work.
DS was supported by EPSRC Early Career research fellowship EP/L020505/1.
\bibliographystyle{ieeetr}
\bibliography{refs}

\ifCLASSOPTIONcaptionsoff
  \newpage
\fi
\end{document}

%% file: abstract.tex
\begin{abstract}
The advance of technology for transmitting Data-over-Sound in various IoT and telecommunication applications has led to the concept of machine-to-machine over-the-air acoustic signalling. Reverberation can have a detrimental effect on such machine-to-machine signals while decoding. Various methods have been studied to combat the effects of reverberation in speech and audio signals, but it is not clear how well they generalise to other sound types. We look at extending these models to facilitate machine-to-machine acoustic signalling.
This research investigates dereverberation techniques to shortlist a single-channel reverberation suppression method through a pilot test.
In order to apply the chosen dereverberation method a novel method of estimating acoustic parameters governing reverberation is proposed. The performance of the final algorithm is evaluated on quality metrics as well as the performance of a real machine-to-machine decoder.
We demonstrate a dramatic reduction in error rate for both audible and ultrasonic signals.
\end{abstract}

%% file: introduction.tex
\section{Introduction}
Digital signal processing (DSP) techniques have led to systems that allow us to use an audio signal as a medium to transfer data among devices. There is a continuous growing demand for transmitting Data-over-Sound for various IoT and telecommunication applications \cite{computingwithsound}\cite{smartappliances}\cite{Nandakumar:2013:DSP:2486001.2486037}\cite{Pearson2013ACQRAQ}\cite{7795155}.
Signals radiated across a room are often corrupted by reverberation. Reverberation is the the dominant acoustic characteristic of a space. The effect of reverberation depends on the physical structure of the room and the objects present in it. At the listener's end, along with the clean signal the superpositions of many delayed and attenuated copies of the clean signal are heard due to multiple reflections from the surroundings. These corruptions degrade the intelligibility of the audio signal.
 Multiple studies have been carried out on the effect of reverberation on speech signals for applications such as Automatic Speech Recognition \cite{pmid22732772}\cite{pmid22356300}\cite{pmid20871182}. There exists a wide range of non-speech digital audio signals being used as a carrier medium, for which it is desirable to remove these reverberant parts to enhance the detection \cite{secondscreen}. However, machine-machine (M2M) signals can be very different from speech, and the decoding requirements are also quite different. It is therefore important to investigate DSP techniques to suppress the distortions caused by an acoustic environment on such data carrying machine-to-machine acoustic signals.
In this research, we extend the work done on existing speech dereverberation models to machine-to-machine acoustic signals.

%% file: conclusion.tex
\section{CONCLUSIONS}
Similar to speech and other digital signals, reverberation is shown to degrade the intelligibility of the source sounds for  decoding of M2M signals. The main goal of this project was to investigate and propose a dereverberation model with suitable parameter estimation to suppress such unwanted influences and therefore increase the decode rate of M2M signals. A pilot study was performed to validate the accurateness of simulating an environment and it is found to be a viable option to test the performance of M2M signals. 
Since Chirp signals fall in the single channel signal, the single channel reverberation suppression techniques were found to be best suited. Spectral subtraction with a novel parameter estimation algorithm is found to be the best dereverberation method for M2M signals. Reverberation parameters required for the dereverberation method are estimated by extending the energy decay algorithm. The evaluation of the model shows that the estimates of the reverberation parameters have a mean error of $\pm0.11s$ for the common real-world environments $(RT_{60} < 2s).$ The overall result of the algorithm combining reverberation estimation model and spectral subtraction show good dereverberation while preserving the packet data. The improvement in decode rate is found to be ~25 percentage points for audible and ~6 percentage points for inaudible Chirp signals. 